\documentclass[useAMS,usenatbib,usegraphicx]{mn2e}
\usepackage{amsmath,amsfonts,amssymb}
\usepackage[latin1]{inputenc}
\usepackage{times} 
\usepackage{color}     
\usepackage{graphicx}
\usepackage{ulem}

\usepackage{epsfig}
\newcommand{\dd}{\textrm{d}}

\def\Msun{\hbox{$\rm\, M_{\odot}$}}
\newcommand{\lgal}{{\sc L-Galaxies}}
\newcommand{\SAM}{SA model}
\newcommand{\spose}[1]{{\hbox to 0pt{#1\hss}}}
\newcommand{\lta}{\mathrel{\spose{\lower 3pt\hbox{$\mathchar"218$}}
     \raise 2.0pt\hbox{$\mathchar"13C$}}}
\newcommand{\gta}{\mathrel{\spose{\lower 3pt\hbox{$\mathchar"218$}}
     \raise 2.0pt\hbox{$\mathchar"13E$}}}
\newcommand{\ta}{t_\mathrm{a}}
\newcommand{\Vespa}{{VESPA}}

\newcommand{\Fig}[1]{Fig.~\ref{fig:#1}}
\newcommand{\Sec}[1]{Section~\ref{sec:#1}}
\newcommand{\Tab}[1]{Table~\ref{tab:#1}}

\title [Star formation histories of galaxies]{Galaxy formation in the {\it
    Planck} cosmology - II.  Star formation histories and post-processing
  magnitude reconstruction}

\author[Shamshiri et al.]
{Sorour Shamshiri$^{1}$,
 Peter A. Thomas$^1$\thanks{E-mail: P.A.Thomas@sussex.ac.uk}, 
 Bruno M. Henriques$^2$, Rita Tojeiro$^3$,\newauthor
 Gerard Lemson$^{2,4}$, Seb J. Oliver$^1$, Stephen Wilkins$^1$\\
 {}$^1$Astronomy Centre, University of Sussex, Falmer, Brighton BN1 9QH,
 UK\\
 {}$^2$Max-Planck-Institut f\"ur Astrophysik,
 Karl-Schwarzschild-Str. 1, D-85741 Garching b. M\"unchen, Germany\\
 {}$^3$School of Physics \& Astronomy, Physical Science Building, North Haugh,
 St.~Andrews, Fife, KY16 9SS, UK\\
 {}$^4$Department of Physics \& Astronomy, Johns Hopkins University, Baltimore,
 MD, 21218, USA\\
}
\begin{document}
\date{MNRAS, in press}

\pagerange{\pageref{firstpage}--\pageref{lastpage}} \pubyear{2013}

\maketitle

\label{firstpage}

\begin{abstract}

We adapt the \lgal\ semi-analytic model to follow the star formation
histories (SFH) of galaxies -- by which we mean a record of the formation time
and metallicities of the stars that are present in each galaxy at a given time.
We use these to construct stellar spectra in post-processing, which offers large
efficiency savings and allows user-defined spectral bands and dust models to be
applied to data stored in the Millennium data repository.

We contrast model SFHs from the Millennium Simulation with observed ones from
the \Vespa\ algorithm as applied to the SDSS-7 catalogue. The overall agreement
is good, with both simulated and SDSS galaxies showing a steeper SFH with
increased stellar mass.  The SFHs of blue and red galaxies, however, show poor
agreement between data and simulations, which may indicate that the termination
of star formation is too abrupt in the models.

The mean star-formation rate (SFR) of model galaxies is well-defined and is
accurately modelled by a double power law at all redshifts:
SFR\,$\propto1/(x^{-1.39}+x^{1.33})$, where $x=(\ta-t)/3.0\,$Gyr, $t$ is the age
of the stars and $\ta$ is the lookback time to the onset of galaxy formation;
above a redshift of unity, this is well approximated by a gamma function:
SFR\,$\propto x^{1.5}e^{-x}$, where $x=(\ta-t)/2.0\,$Gyr.  Individual galaxies,
however, show a wide dispersion about this mean.  When split by mass, the SFR
peaks earlier for high-mass galaxies than for lower-mass ones, and we interpret
this downsizing as a mass-dependence in the evolution of the quenched fraction:
the SFHs of star-forming galaxies show only a weak mass dependence.
\end{abstract}

\begin{keywords}
methods: numerical -- galaxies: formation -- galaxies: evolution
\end{keywords}

\section{Introduction}
\label{sec:intro}

Understanding the astrophysics behind the formation and evolution of galaxies is
an important goal in modern astronomy.  One of the most fundamental probes of
that physics is the star formation rate as a function of cosmic time.  In this
paper we contrast predicted and observed star formation histories (hereafter,
SFHs) of galaxies, and explore the expected range of SFHs at high redshift.

Two main observational approaches are used to infer the SFH
of galaxies. One
can look at the instantaneous star formation rate as a function of cosmic time
\citep[for an overview see, e.g.~][]{Ken98,Cal99}, or one can deduce the history
from the fossil record of current-day galaxies.  The two techniques in fact
measure slightly different things, with the relationship between the two
depending upon the merging history of the galaxies.  

This paper focuses upon the archaeological approach. We have extended the
\lgal\ semi-analytic model (hereafter simply \lgal) to keep a record of the SFHs
of individual galaxies, with a bin-size that increases with increasing age for
the stars.  The resulting data have been made available as part of the public
data release (DR) of the Millennium Simulation\footnote{\tt
  http://gavo.mpa-garching.mpg.de/MyMillennium/} \citep{Lem06} that accompanies the latest
implementation of \lgal\ \citep{HWT14}.

We note that the SFH as defined in this paper (the history of star formation of
all the stars that end up in the galaxy at some particular time) is related to,
but distinct from, the variation in mass of a galaxy over time.  The latter
follows only the history of the main component of the galaxy (along the ``main
branch'' of the merger tree) and has been investigated for the Millennium
Simulation by \citet{CoV14}.  The difference between the two reflects the merger
history of galaxies.

The term SFH is often loosely used in papers without being defined.
Observationally, the only {\it direct} measure of SFHs corresponds to that
described in this paper, i.e.~the distribution of formation times of all the
stars that make up the galaxy.\footnote{We note that this would better be
  described by the phrase ``stellar age distribution'' rather than
  star formation history, but the latter phrase predominates in the literature.}  
Measures of the SFH of the main galactic
component can only be inferred statistically by observing populations of
galaxies at different redshifts and making some assumptions about merger rates
as a function of stellar mass and environment.  In principle, the former method
is much cleaner, as it is free from these model assumptions; however, in
practice, the inversion of the stellar spectra is highly degenerate and
sensitive to the input stellar population synthesis models, and can lead to
implausible results if some model constraints are not imposed.

As well as enabling comparison to observations, the introduction of SFHs into
\lgal\ allows reconstruction of galaxy magnitudes using arbitrary stellar
population synthesis and dust models, in any band, in post-processing, and we
investigate the accuracy of this approach.  In addition, having SFHs is a
prerequisite for a correct, time-resolved treatment of galactic chemical
enrichment, as described in \citet{YHT13}.

The star formation history of a galaxy, its chemical evolution, and its current dust
content, can in principle be fully recovered with high enough quality
observational data, suitable modelling of the spectral energy distribution of
stellar populations and dust extinction, and appropriate
parametrization. Several algorithms have been developed over the last decade
that attempt to do the above in the most robust way (see e.g. {\sc moped} by
\citealt{HPJ04, PJH07}, {\sc stecmap} by \citealt{OPLT06}, {\sc starlight} by \citealt{CF04,
  CF05} or {\sc ulyss} by \citealt{KPBW09}). In this paper we focus on the results
obtained by VESPA \citep{THJ07}, a full spectral fitting code that was applied to
over 800,000 Sloan Digital Sky Survey DR7 galaxies \citep{SDSS7}. The
resulting data base of individual SFHs, metallicity histories
and dust content is publicly available and described in \citet[hereafter
  TWH09]{TWH09}. The wide range of galaxies in SDSS DR7 and the time resolution
of the SFHs published in the data base make it ideal for a detailed comparison
with our model predictions.

At higher redshift, observational data are scarce.  Nevertheless, we make
predictions from our models of how the star formation rate of galaxies evolves
with redshift, and we show how downsizing arises from a mass-dependence of the
rate at which galaxies are quenched.

In outline, the main aims of this paper are:
\begin{itemize}
\item to briefly overview the \lgal\ and \Vespa\ algorithms --
  Sections~\ref{sec:lgal} and \ref{sec:vespablurb};
\item to describe the \lgal\ SFH binning method -- \Sec{sfhbin};
\item to test how well post-processing can reconstruct magnitudes --
  \Sec{ppmag};
\item to compare model galaxies from \lgal\ with
  results from the \Vespa\ catalogue -- \Sec{vespa};
\item to investigate the variety of SFHs that we find in our
  model galaxies -- \Sec{sfh};
\item to provide a summary of our key results -- \Sec{conc}.
\end{itemize}

\section{Methods}
\label{sec:methods}

\subsection{\lgal}
\label{sec:lgal}

In this work we use the latest version of the Munich semi-analytic
  (SA) model, \lgal, as described in \citet{HWT14}.  This gives a good fit to,
amongst other things, the observed evolution of the mass and luminosity
functions of galaxies, the fraction of quenched galaxies, the star formation versus
stellar mass relation (at least at $z<2$), the Tully--Fisher relation,
metallicities, black hole masses, etc. We refer to this standard model as
HWT14. The improvements obtained in terms of the evolution of the abundance and
red fractions of galaxies as a function of stellar mass in this model result
from: a supernova feedback model in which ejected gas is allowed to fall back
on to the galaxy on a time-scale that scales inversely with halo virial mass
\citep{HWT13}; and a lower star formation threshold and weaker environmental
effects both reducing the suppression of star formation in dwarf galaxies.

\lgal, as with other \SAM{}s, follows the growth of galaxies within the
framework of a merger tree of dark matter haloes.  We construct this tree from
the Millennium Simulation \citep{SWJ05}, scaled using the method of
\citet{AnW10} to the {\it Planck} cosmology \citep{Planck13}:
$\Omega_\mathrm{m}=0.315$, $\Omega_\Lambda=0.683$, $\Omega_\mathrm{b}=0.0488$,
$n_\mathrm{s}=0.958$, $\sigma_8=0.826$, $h=0.673$.  This then gives a box size of
$480.3\,h^{-1}$\,Mpc and a particle mass of $9.61\times10^{8}h^{-1}$\,\Msun.
The merger tree is constructed from 58 snapshots,\footnote{Five of the original
  {\it Millennium Simulation} snapshots lie at $z<0$ after scaling.}  each of
which is subdivided into 20 integration timesteps.  The snapshots are unevenly
spaced, such that the time resolution is higher at high redshift, but a typical
timestep is of the order of 1--2$\times10^7$\,yr. We note that the \SAM\ has been
implemented on both the higher resolution Millennium-II \citep{BSW09} and larger
volume Millennium-XXL simulations \citep{AWS14} although we do not make use of
those simulations in the current work.

Prior to HWT14, galaxy luminosities were determined by accumulating flux in
different spectral bands throughout the time-evolution of each galaxy.  In the
new model those fluxes are recovered to high accuracy in post-processing, by
simply recording the star formation and metallicity history in a relatively
small number of time-bins. The new method is introduced following the work
outlined and presented in the current paper. The DR that accompanies this series
of papers records the SFHs, allowing the user flexibility to define their own
bands and dust models.

\subsection{\Vespa}
\label{sec:vespablurb}

The spectrum of galaxy, in the absence of dust, can be described as the linear
superposition of the spectra of the stellar populations of different ages and
metallicities that exist in the galaxy. The deconvolution of a galaxy's spectrum
into a star formation and metallicity history is in principle trivial, but
complicated by noisy or incomplete data and limitations in the
modelling. \cite{OPLT06} showed how the problem quickly becomes ill conditioned
as noise increases in the data, and that the risk of overparametrizing a galaxy
is high. VESPA takes into account the noise and data quality of each individual
galaxy and uses an algebraic approach to estimate how many linearly
independent components one can extract from each observed spectrum, thereby
avoiding fitting the noise rather than the signal - see \cite{THJ07} for details.

VESPA recovers the SFH of a galaxy in 3 to 16 age bins
(depending on the quality of the spectra), logarithmically spaced between 0.002
Gyr and the age of the Universe. For each age bin, VESPA returns the total mass
formed within the bin and a mass-weighted metallicity of the bin, together with
an estimate of the dust content of the galaxy.  As we always compare our
model predictions to the mean SFH of large ensembles of galaxies, we choose to
use the fully-resolved SFHs published in the data base of TWH09. Whereas we
expect these to be dominated by the noise on each individual galaxy, the mean
over a large ensemble has been shown to be robust \citep{PHJ03}.  In this paper,
we will compare to mean SFHs for different galaxy
populations, as described in the text.

\subsection{The SFH binning algorithm}
\label{sec:sfhbin}

\begin{figure}
\centering
\includegraphics[width=8.4cm]{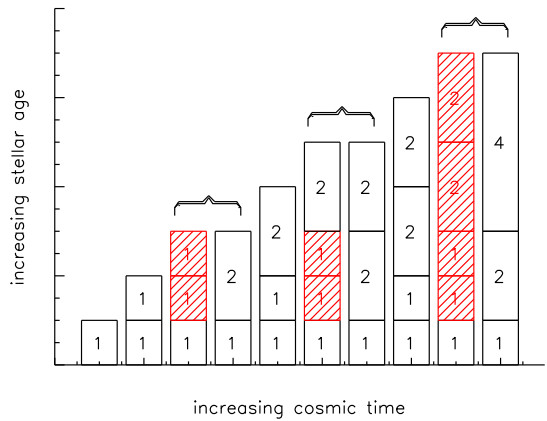}
\caption{The evolution of the SFH bins for $N_\mathrm{max}=2$.  The
  $x$-axis represents successive timesteps, starting from high redshift and
  moving towards the present day.  The $y$-axis represents bins of stellar age,
  starting with stars that are newly-created at that time and looking back into
  the past.  The numbers within each bin represent the number of timesteps that
  have been merged together to produce that bin.  The columns that are bracketed
  together show different arrangements of the data at a single cosmic time.  The
  shaded, red stacks represent transient structures in which some bins merge
  together to produce the black stacks to their right.}
\label{fig:sfhbins}
\end{figure}

As mentioned in Section~\ref{sec:lgal}, the Millennium merger trees
are constructed from 58 snapshots, each of which is separated into 20
integration timesteps.  To follow the history of star formation,
we introduce extra arrays to carry information on the mass and
metallicity of stars in each component of the galaxy (disc, bulge,
intracluster mass) as a function of cosmic time.  To save the
information over all 1160 timesteps would consume too much memory and
is unnecessary.  Instead, we wish to use a high resolution for the
recent past (when the stellar population is rapidly evolving) and a
lower one at more distant times.  To do so, we adopt the following
procedure (see \Fig{sfhbins}).

Starting at high redshift, on each timestep a new bin is created to hold the
SFH information.  Whenever the number of bins of a particular
resolution exceeds $N_\mathrm{max}$ (where in the diagram, for the purposes of
illustration, $N_\mathrm{max}=2$), then the two oldest bins are merged together
to form a new bin of twice the size -- this may result in a cascade of mergers
at successively higher levels (in the figure, these mergers are represented by
red columns joined by braces to the merger product).  In this way, the number of
bins at each size grows from 1 to $N_\mathrm{max}$, then oscillates between
$N_\mathrm{max}$ and $N_\mathrm{max}-1$.  The total number of bins required does
not exceed the smallest integer greater than
$N_\mathrm{max}\log_2(N_\mathrm{step}/N_\mathrm{max}+1)$, where
$N_\mathrm{step}$ is the number of timesteps.  \Tab{sfhbin} shows, for the {\it
  Millennium Simulation}, using 20 steps within each of 58 snapshots, the
maximum number of bins required, and, at $z=0$, the actual number of SFH bins
and their minimum and maximum size in years.\footnote{These numbers are
  unchanged for the 63 snapshots required to use the {\it Millennium Simulation}
  with the original {\it WMAP}-1 cosmology.}  Note that all choices of
$N_\mathrm{max}\geq2$ have the same minimum bin-size, equal to that of the
original timesteps; what differs is the number of bins that are resolved at that
highest resolution.

\begin{table}
\begin{center}
\caption{For the {\it Millennium Simulation}, using 1160 timesteps, this table
  shows, for different choices of $N_\mathrm{max}$: $N_\mathrm{tot}$ -- the
  maximum number of SFH bins required; $N_{z=0}$ -- the number of populated bins
  at $z=0$; $\Delta t_\mathrm{min}$/yr -- the minimum bin-size in years at
  $z=0$; $\Delta t_\mathrm{max}$/yr -- the maximum bin-size in years at
  $z=0$.}
\label{tab:sfhbin}
\begin{tabular}{ccccc}
\hline
$N_\mathrm{max}$& $N_\mathrm{tot}$& $N_{z=0}$& $\Delta
t_\mathrm{min}$/yr& $\Delta t_\mathrm{max}$/yr\\
\hline
1& 11& 7& $6.0\times10^7$& $1.1\times10^{10}$\\
2& 19& 16& $1.5\times10^7$& $2.1\times10^9$\\
3& 27& 23& $1.5\times10^7$& $1.6\times10^9$\\
4& 34& 31& $1.5\times10^7$& $5.6\times10^8$\\
\hline
\end{tabular}
\end{center}
\end{table}

We have investigated the sensitivity of our results to the number of bins and
conclude that $N_\mathrm{max}=2$ gives the best balance between data-size and
accuracy: that is the value used in the Millennium data base and, unless
mentioned otherwise, in the results presented below.

\begin{table*}
\begin{center}
\caption{This table shows the root-mean-square difference between magnitudes
  calculated on-the-fly during the running of the code and those calculated in
  post-processing.}
\label{tab:ppmag}
\begin{tabular}{lcccccc}
\hline
& & Without dust  & & & With dust &\\
$z=0$ & \multicolumn{3}{c}{\hrulefill}& \multicolumn{3}{c}{\hrulefill}\\
&$N_\mathrm{max}=1$ &$N_\mathrm{max}=2$
&$N_\mathrm{max}=4$  &$N_\mathrm{max}=1$
&$N_\mathrm{max}=2$ &$N_\mathrm{max}=4$ \\
\hline
{\it GALEX} $FUV$               &3.02 &0.29 &0.16 &2.53 &0.38 &0.29 \\
{\it GALEX} $NUV$              &1.79 &0.05 &0.03 &1.47 &0.16 &0.16 \\
SDSS $u$                     &0.76 &0.03 &0.01 &0.66 &0.06 &0.05 \\
SDSS $g$                     &0.50 &0.02 &0.00 &0.46 &0.03 &0.02 \\
SDSS $z$                     &0.36 &0.02 &0.00 &0.33 &0.02 &0.02 \\ 
VISTA $J$                     &0.35 &0.02 &0.00 &0.32 &0.02 &0.02 \\ 
VISTA $K_s$                 &0.33 &0.01 &0.01 &0.31 &0.02 &0.02 \\ 
IRAC $3.6_{\mu \rm{m}}$ &0.38 &0.01 &0.00 &0.36 &0.03 &0.02 \\
\hline
& & without dust  & & & with dust &\\
$z=2$ & \multicolumn{3}{c}{\hrulefill}& \multicolumn{3}{c}{\hrulefill}\\
&$N_\mathrm{max}=1$ &$N_\mathrm{max}=2$
&$N_\mathrm{max}=4$  &$N_\mathrm{max}=1$
&$N_\mathrm{max}=2$ &$N_\mathrm{max}=4$ \\
\hline
{\it GALEX} $FUV$               &1.74 &0.04 &0.01 &1.14 &0.27 &0.26 \\
{\it GALEX} $NUV$              &1.39 &0.02 &0.00 &0.95 &0.21 &0.20 \\
SDSS $u$                     &0.88 &0.02 &0.00 &0.63 &0.13 &0.12 \\
SDSS $g$                     &0.64 &0.01 &0.00 &0.47 &0.09 &0.09 \\
SDSS $z$                     &0.39 &0.01 &0.00 &0.33 &0.04 &0.04 \\ 
VISTA $J$                     &0.33 &0.01 &0.00 &0.30 &0.02 &0.02 \\ 
VISTA $K_s$                 &0.32 &0.01 &0.00 &0.30 &0.01 &0.01 \\ 
IRAC $3.6\mu$m &0.31 &0.01 &0.00 &0.28 &0.02 &0.02 \\
\hline
\end{tabular}
\end{center}
\end{table*}

\begin{figure*}
\centering
\includegraphics[width=8.6cm]{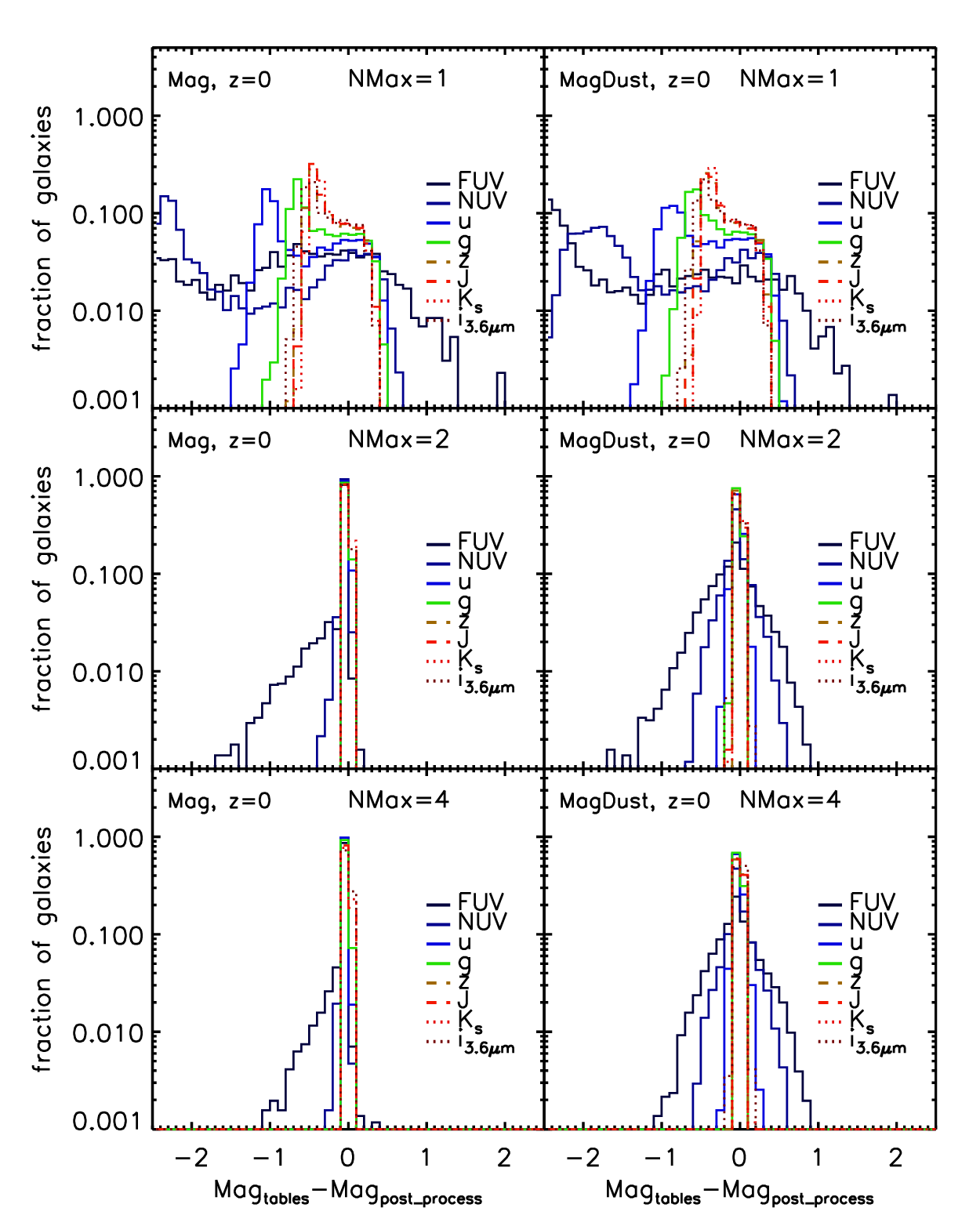}
\hspace{0.1cm}
\includegraphics[width=8.6cm]{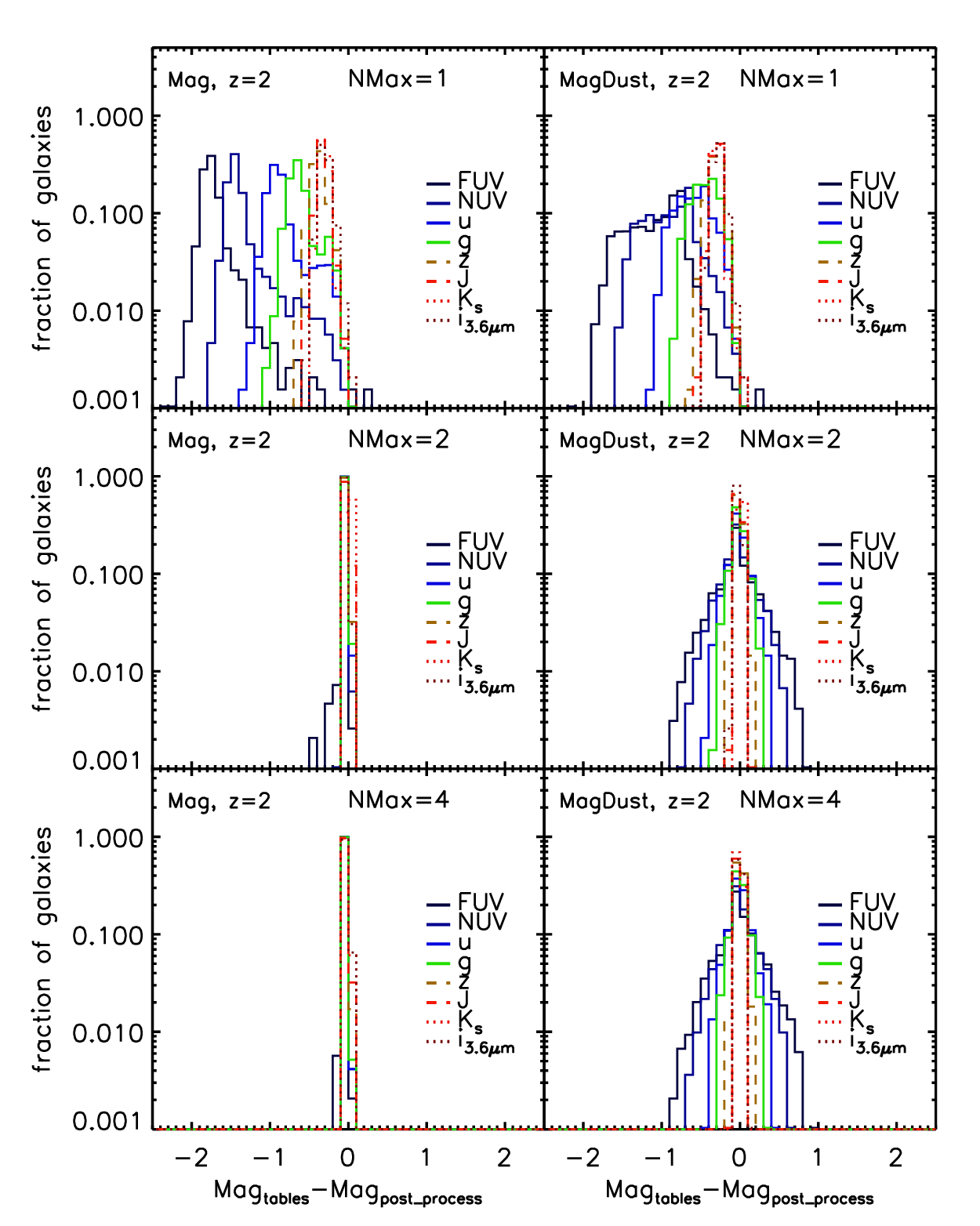}
\caption{The difference between photometric properties calculated using full
  resolution and binned SFHs at $z=0$ (left) and $z=2$
  (right). The left-hand column in each set of panels correspond to intrinsic
  magnitudes while the right takes into account dust extinction. From top to
  bottom the resolution of the binning is increased: $N_\mathrm{max}=1$ (top), 2
  (middle) and 4 (bottom).}
\label{fig:ppmags}
\end{figure*}

\section{Post-processing of magnitudes}
\label{sec:ppmag}

In most \SAM{}s, and in \lgal\ prior to this
work, galaxy luminosities are computed by adding the flux in different
bands throughout the time-evolution of each galaxy. This calculation
generally requires interpolating between values in large stellar
population synthesis tables and represents a large fraction, and in
some cases the majority, of the computational time for the entire
galaxy formation model. The problem is aggravated as different types
of magnitudes (dust corrected, observer-frame) for additional
components (e.g.~disc, bulge, intra-cluster light) are included.

These difficulties can, in principle, be circumvented by storing the star
formation and metallicity histories for different components of the galaxies,
and using them to compute emission in post-processing. Ideally, this history
would be stored for all the intermediate steps between output snapshots for
which galaxy properties are computed. However, memory constraints make this
infeasible. For our current set up, for example, it would require storage of
up to 2320 values for each galaxy component (58 snapshots, 20
intermediate steps per snapshot, star formation and metallicity).

An alternative, tested in this paper, is to store the histories in bins that
grow in size for older populations, as described in \Sec{sfhbin}. Since the
emission properties of populations vary on significantly longer time scales for
old populations this can in principle allow us to maintain accuracy. To validate
the method we compare the theoretical emission from galaxies computed both with
full resolution on-the-fly and from star formation and metallicity history bins
that are merged together for older populations. We assume that star formation
occurs at a time corresponding to the mid-point of each SFH bin.  To spread
star formation out over the time-bin would be equivalent to using a larger
number of timesteps (which we have also tested) and makes little difference
except in the UV.

\Fig{ppmags} shows the difference between photometric properties calculated
using full resolution and binned SFHs for $z=0$ and
$z=2$. In both sets of panels the left-hand column corresponds to intrinsic
magnitudes while the right-hand column takes into account dust extinction. From top
to bottom the resolution of the binning is increased: $N_\mathrm{max}=1$, 2 and
4. While large differences between the two methods are seen for the lowest
resolution, the figures show that good convergence is achieved for
$N_\mathrm{max}=2$ or more.

Table~\ref{tab:ppmag} shows the root-mean-square (rms) difference between magnitudes
calculated on-the-fly during the running of the code (i.e.~using the finest
possible time resolution) and those calculated in post-processing, using
different numbers of SFH bins.  Quantitatively, for $N_\mathrm{max}=2$, at
$z=0$, the rms difference between the two methods is less than 0.05 for all
intrinsic magnitudes except the far-UV, for which it is approximately
0.29.\footnote{These values are reduced to 0.03 and 0.16, respectively, when
  $N_\mathrm{max}=4$ is used.}  At $z=2$, the mean difference is less than 0.04
in all bands.

The increased accuracy in the far-UV at high redshift results from
the higher accuracy in post-processing at a time in which fewer bins were
merged.  Emission in this part of the spectrum is dominated by extremely young
populations for which even a slightly different formation time results in a
large variation in predicted flux.  The two methods thus differ in detail, but
have the same statistical properties.  If one is interested in the precise UV
flux of a particular galaxy then that can be recovered by using finer time-bins.
We have checked that, keeping $N_\mathrm{max}=2$ but using a finer timestep
(which adds very few SFH bins) improves the agreement between the two methods of
calculating fluxes.\footnote{We do not use a finer time resolution as a default
  as this agreement is illusory -- the underlying merger tree is not capturing
  the dynamics on that short a timescale.}

At low redshift, it can be seen that there is a residual error in the calculated
intrinsic far-UV flux even for $N_\mathrm{max}=4$.  This arises from stars of
age about 1\,Gyr (i.e. the TP-AGB population), for which the tabulated UV fluxes
in the \citet{Mar05} population synthesis tables show a large jump in luminosity
between the two lowest-metallicity bins.  Merging galaxies that contain stars of
differing metallicity can therefore lead to large changes in flux.  Such merging
can occur for any choice of $N_\mathrm{max}$ and is hence a fundamental (albeit
very minor) limitation of the SFH magnitude-reconstruction method.\footnote{The
  limitation could be overcome by keeping SFHs for several different metallicity
  bins, but this moves away from the spirit of the method and we do not think
  that the gain justifies the extra storage cost.}

The addition of dust significantly degrades the agreement between the two
reconstruction methods in all bands, with the far- and near-UV being most
affected.  At $z=2$ the rms differences between dust-corrected magnitudes are
approximately 0.20 for far- and near-UV, 0.10 for $u$ and $g$, 0.04 for $z$
and $J$, and 0.02 for $K_s$ and $\rm{irac}_{3.6\mu m}$.  

In the current version of \lgal\, a two-component dust model applies extinction
separately from the diffuse interstellar medium and from molecular birth clouds
(see Section 1.14 in the supplementary material of \citealt{HWT14} for details).
The large differences seen for dust corrected magnitudes are mostly caused by
the latter.  The calculation of this optical depth includes a random gaussian
term that leads to differences in the amount extinction assumed for each
individual galaxy when a different number of timesteps are used.

The method successfully tested in this section is adopted in the recent major
release of the Munich model, HWT14. By computing emission properties in
post-processing, the memory consumption of the code is no longer dependent on
the number of photometric bands.  Moreover, the method allows emission
  properties to be computed after the model is completed using any stellar
  populations synthesis code and for the filters used by any observational
  instrument. To show the potential of the new method, the new major release
already includes emission in 20 bands in the snapshot catalogues, and in 40
bands and for two different stellar populations for the lightcones, all
calculated in post-processing.

\section{Comparison of \Vespa\ and \lgal}
\label{sec:vespa}

For reasons described in Appendix~\ref{app:vespa}, we use the version of the
\Vespa\ catalogue that was created using the population synthesis model of
\citet{Mar05} with a one-component dust model.

Note that the \Vespa\ data and the \SAM{}s produce output with very
different binning.  The time resolution in the observations is necessarily very
coarse at high redshift, whereas there is no such restriction in the models.  In
\Sec{mgs} below, we re-bin the model predictions to match those of \Vespa;
throughout the rest of the paper, we will keep the actual binning returned by
the models so as to allow a clearer understanding of the growth of galaxies at
high redshift.

\subsection{The main galaxy sample}
\label{sec:mgs}

\begin{figure}
\centering
\includegraphics[width=85mm]{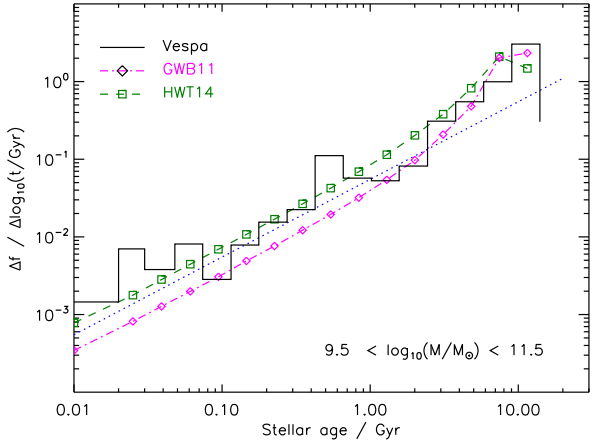}
\caption{The average SFH from \Vespa\ (black lines), the
  GWB11 model (magenta, dash-dotted lines) and the HWT14 model (dashed, green
  lines) within the indicated mass range.  To guide the eye,
  the blue, dotted line is the same in each panel and has slope unity,
  corresponding to a constant SFR.}
\label{fig:all}
\end{figure}

The main SDSS galaxy sample covers a redshift range of $0<z\lta0.35$.  In
Fig.~\ref{fig:all} we show the mean SFH for all galaxies using the maximum
\Vespa\ resolution of 16 bins. For most galaxies, the data quality is not good
enough to independently measure masses in all 16 bins and so the
\Vespa\ algorithm will return solutions on bins of varying and lower-resolution
width, as described in TWH09.  The assumed star formation rate (hereafter, SFR)
within each bin depends on its width (it is constant in high-resolution bins,
and exponentially decaying in low-resolution bins). Choosing a SFR within a bin
is an unavoidable part of the process of parametrizing a galaxy. We have however
checked that our conclusions remain unchanged if we: (i) use only galaxies with
high-resolution bins, and (ii) use a constant SFR in wide bins.
  
For each \Vespa\ galaxy, we select the model galaxy that most closely matches it in
mass and redshift, then use this to construct a mean SFH.
The result is shown in \Fig{all} along with the predictions from two versions of
\lgal: GWB11--\citet[magenta diamonds and dot--dashed line]{GWB11} and
HWT14--\citet[green squares and dashed lines]{HWT14}. 

In this and subsequent plots, $\Delta f$ is defined as the fraction of stars
(i.e.~the specific stellar mass) within each bin.  Unless mentioned otherwise,
$\Delta f$ is calculated separately for each galaxy and then averaged
(i.e.~weighted by galaxy number rather than stellar mass).

The \Vespa\ results are not as smooth as the \SAM\ data. We expect the averaged
rest-frame SFH of a varied ensemble of galaxies to be devoid of significant
structure, and to appear smooth as seen on the model galaxies. At late times,
especially, the SFR should be approximately constant, and so one would expect
the SFH bins to run parallel to the dotted line, of slope equal to unity. The
features seen on the data are heavily dependent on the modelling (see
e.g. TWH09, \citealt{Toj13}, Appendix~\ref{app:vespa}), and therefore are the
likely result of limitations of the stellar population synthesis and dust
modelling.

The \SAM{}s show a turn-down in SFR in the oldest stellar age bin, corresponding
to the onset of star formation.  No such feature is seen in the \Vespa\ results,
most probably because the spectral signatures are too weak to be detected and so
the reconstruction method imposes a declining SFR even at these earliest times.

Other than that, at ages above 1\,Gyr the GWB11 model seems to provide a
reasonable match the observations, whereas for younger stars, the HWT14 model
is a better fit.  To draw more definitive conclusions about which is the
preferred model, one would have to look in much more detail at the
reconstruction biases that may be present in the \Vespa\ method when applied to
imperfect data, and that will be the subject of future work.

From here on we will rebin the VESPA solutions to five bins in age, to reduce the
scatter and to average over these features that we know to be unphysical.

\subsection{Mass selection}
\label{sec:mass}

\begin{figure}
\centering
\includegraphics[width=85mm]{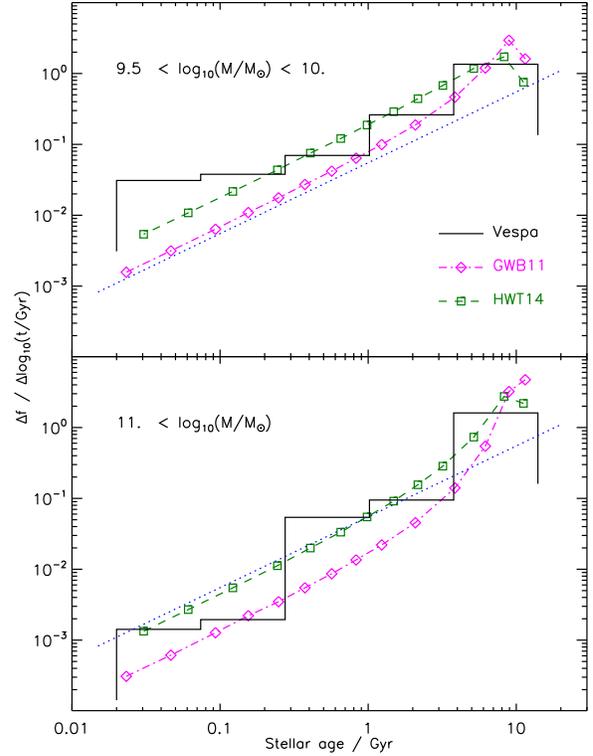}
\caption{The average SFH from \Vespa\ (black lines), the
  GWB11 model (magenta, dash-dotted lines) and the HWT14 model (dashed, green
  lines) within two different stellar mass bins, as shown.  To guide the eye,
  the blue, dotted line is the same in each panel and has slope unity,
  corresponding to a constant SFR.}
\label{fig:mass}
\end{figure} 

\Fig{mass} shows the SFHs broken down by stellar mass. The
\Vespa\ reconstruction gives a slope that is too steep for high-mass galaxies
and too shallow for low-mass galaxies, when compared to the expected constant
SFR at recent times. Nevertheless, it can be seen that both the
\Vespa\ galaxies, and those in the \SAM, form stars earlier in higher-mass
galaxies and have a correspondingly lower SFR at late times.

The variation of the model SFHs with mass is explored further in \Sec{sfhmass},
below.

\subsection{Colour selection}
\label{sec:colour}

\begin{figure}
\centering
\includegraphics[width=85mm]{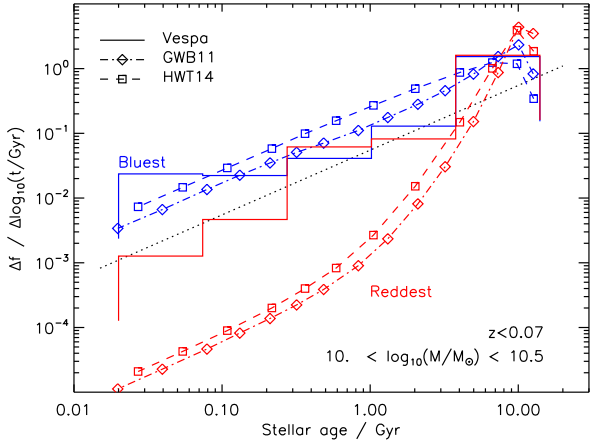}
\caption{The averaged SFH from VESPA (solid lines),
  the GWB11 models (dashed-dotted lines), and the HWT14 model (dashed
  lines) for red and blue galaxies with masses greater than $10^{11}$\,\Msun and
  within redshift interval $0.10<z<0.12$.}
\label{fig:colour}
\end{figure} 

Next, we look at the distinction between red and blue galaxies by selecting
according to $u-r$ colour.  As HW14 showed, in spite of reproducing the observed
galaxy colour bimodality, \lgal\ does not reproduce the exact colour
distributions seen in SDSS. Therefore, applying the same colour cuts in the data
and simulation would result in picking out intrinsically different galaxy
populations. Instead, we select the 10\% bluest and reddest galaxies, according
to $u-r$ colour, in both the \Vespa\ and \lgal\ samples.  The resulting SFHs are
show in \Fig{colour} for galaxies of mass $10^{10}<M/\Msun<10^{10.5}$ at low
redshift, $z<0.07$.

VESPA produces similar SFHs for both blue and red galaxies except for the
youngest stars of age less than $3\times10^8$\,yr.  At first sight, it seems
surprising that the deviation between the two populations can have occurred so
recently.  One interpretation is that galaxies in this mass range may transition
back and forth between star forming and quiescent (i.e.~show bursts of star
formation) on time-scales of this order, and that might also help to explain why
the SFR of the bluest galaxies seems to increase to the present day.  However,
this seems at odds with the observation that red and blue galaxies are observed
to have very different metallicities: the stellar-mass-weighted metallicity of
the young (age less than 2.5\,Gyr) stars is 0.036 in red galaxies,\footnote{In
  units of the mass fraction of metals with respect to Hydrogen; in these units
  solar metallicity is $Z_\odot = 0.02$} and 0.019 in blue galaxies, which would
suggest that the two form distinct populations.

Whatever the interpretation of the observations, it seems unlikely that they
can be made compatible with the model galaxies, which show widely divergent SFHs
for red and blue galaxies for stars younger than 5\,Gyr.  It would seem that
termination of star formation is too abrupt in the models as compared the
observations, and lacks the possibility of retriggering of star formation at
later times.

The comparison of observed and model galaxies is complicated by the effects of
metallicity, dust attenuation and finite fibre aperture on the measured colours.
In the models, there is a very strong correlation between stellar mass and
metallicity, and between SFR and extinction; in addition; there
is no aperture correction.  In real galaxies, the scatter is observed to be much
higher, and there will be a redshift-dependent colour correction for the finite
aperture.  This, together with observational error, is likely to move the
observed blue and red populations towards each other, so caution should be
exercised before drawing definitive conclusions. On the other hand, \SAM{}s
struggle to match even the colour distribution of galaxies (see, for example,
Fig. 9 of HWT14), and the distinction between the models and the observed SDSS
data is so large that it is hard to dismiss it lightly.  This issue will be
investigated in a subsequent paper, and highlights the power in comparing
fully-resolved SFHs between models and data.

\section{The evolution of SFHs}
\label{sec:sfh}

This section looks at the predicted SFHs of galaxies at different redshifts.  We
are interested both in the history of the mean (and median) galaxy population
and of the scatter about that mean.  This can have important implications for
the interpretation of high-redshift galaxies that often rely upon postulated
SFHs \citep{BBP14,PCC15}.

\subsection{Mean SFHs}
\label{sec:sfhmean}

\begin{figure}
\centering
\includegraphics[width=85mm]{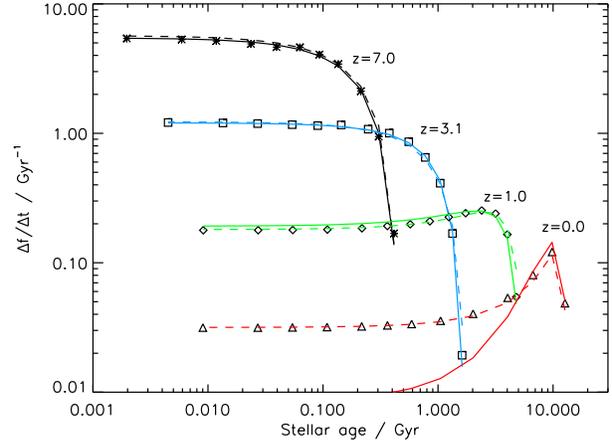}
\caption{The average star formation rates of model galaxies from the HWT14
  model, with mass greater than 10$^{9.5}$\Msun, at four different redshifts,
  as shown.  The symbols show the model predictions; the curves are
  fits to the data as described in the text: solid lines, gamma model
  (Equation~\ref{eq:sfhgamma}); dashed lines, two-power model
  (Equation~\ref{eq:sfhtwop}).}
\label{fig:sfhmean}
\end{figure} 

\Fig{sfhmean}\ shows the mean, mass-weighted SFRs of all galaxies with mass
greater than 10$^{9.5}$\Msun\ at four different redshifts: 0, 1, 3 and 7.  (Note
that this plot differs from previous ones in that we are plotting $\Delta
f$/$\Delta t$ rather than $\Delta f$/$\Delta \log t$; we do this to make it
easier to detect any decrease in the SFR at recent times).  The SFR increases
rapidly at early times, then slows down, with a decline to late times (i.e.~low
stellar ages) being apparent for $z\lta 2$.

To characterize the SFR at high redshift, we fit
a gamma model featuring a power-law increase in star formation at early times
followed by an exponential decline:\footnote{The same functional form with $p=1$
  was shown by \citet{SWC14} to be a good fit to the individual SFHs of most of
  their galaxies in SPH simulations of galaxy formation.}
\begin{equation}
{\dd f\over\dd t}=Ax^pe^{-x},\ \ \ x={\ta-t\over\tau}.
\label{eq:sfhgamma}
\end{equation}
Here $\ta$ is the age of the galaxy, $p$ sets the rate at which star formation
builds up, and $\tau$ is the characteristic time-scale over which star formation
declines.  

At all redshifts above $z=1$, the SFHs are well-fit by a single set of
parameters, $p=1.5$ and $\tau=2.0$\,Gyr, with only $\ta$ varying to reflect the
age of the galaxy.\footnote{For $z\gta2$ these parameters are degenerate, but we
  choose to freeze them at the values found at lower redshift.}  That this is
the case is not surprising but reflects that fact that the majority of stars in
the Universe are born within galaxies whose mass exceeds $3\times10^9$\,\Msun:
each of the SFHs shown in \Fig{sfhmean} then mirrors the cosmic SFH.  Using this
set of parameters then star formation begins in our model at $z\approx12$,
0.4\,Gyr after the big bang, and levels off (i.e.~d$^2f$/d$t^2=0$) 3\,Gyr later,
at $z\approx2$.

At lower redshifts, it becomes apparent that an exponential decline is too
steep.  Instead, a two-power model is preferred:
\begin{equation}
{\dd f\over\dd t}={A\over x^{-p}+x^q},\ \ \ x={\ta-t\over\tau}.
\label{eq:sfhtwop}
\end{equation}
Taking $p=1.39$, $q=1.33$ and $\tau=3.0$\,Gyr gives a good fit at all redshifts.

\citet{BWC12} found identical fitting formulae to those used here to be good
fits to the SFHs of galaxies in their abundance matching method to populate
haloes with galaxies that match observed stellar mass functions and
SFRs.  However, for $10^{12}$\,\Msun haloes, they find a value of
$q$, that determines the rate of decay of the SFR at late times, to be
significantly higher than that quoted above: the reason for this difference is
not clear.

We stress that the curves shown in \Fig{sfhmean} are for the mean star formation
rate averaged over all galaxies with mass greater than $10^{9.5}$\Msun.  As is
apparent from \Fig{mass}, high mass galaxies form their stars earlier, and low
mass galaxies later, than the mean trend.  We show in \Sec{sfhmass} that
this is driven primarily by a mass-dependence in the cessation of
star formation, and in \Sec{sfhind} that there is considerable variation
between individual galaxies.

\subsection{Specific SFRs and quiescent fractions}
\label{sec:ssfr}

\begin{figure}
\centering
\includegraphics[width=85mm]{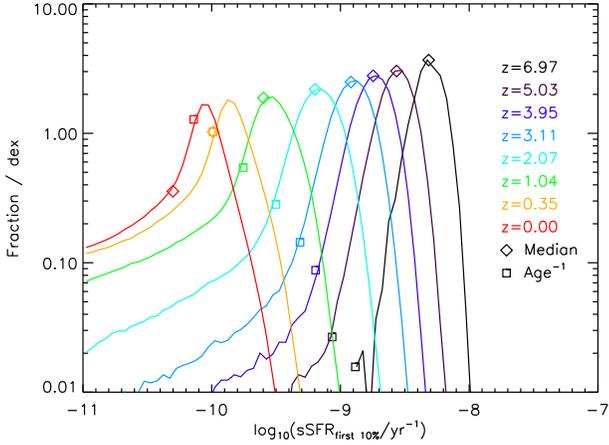}
\caption{The sSFR at different redshifts for the HWT14
  model. The subscript ``first\ 10\,\%'' refers to an average over the most
  recent 10 per cent of the age of the Universe at that time.  The diamonds show
  the median values and the squares show the inverse age of the Universe at each
  redshift.}
\label{fig:ssfr}
\end{figure} 

\Fig{ssfr} shows the specific SFR (sSFR) of all galaxies with
$M>3\times10^{9}$\,\Msun\ in the HWT14 mode at several different redshifts.  The
SFR here is averaged over the most recent 10 per cent of the age
of the Universe at that time.  The diamond symbols show the location of the
median values, and the squares show the inverse age of the Universe at that
redshift.

There is a small spread around the modal sSFR, with the
vast majority of star forming galaxies lying within about $\pm0.25$\,dex of the
peak.  However, there is a long tail of galaxies extending to low sSFRs, which
becomes more prominent at low redshifts; indeed, at $z=0$, 35\,per cent lie off the
left-hand edge of the plot altogether.  That is why, below $z=1$, the median
values lie well to the left of the mode.

Above $z\approx0.35$, most galaxies are forming stars at a rate that would more
than double their mass in the age of the Universe;\footnote{Note that this is a
  number-weighted average, so that does not mean that mean SFR
  peaked at that time.} below that redshift, the opposite is true.  It is at
this time that there is a strong shift from star forming to non-star forming
galaxies.  There is no sharp distinction between the two but, using
\Fig{ssfr} as a guide, we define a galaxy to be {\it quiescent} if it
has formed fewer than 3 per cent of its stars in the most recent 10 per cent of
the age of the Universe, $t_z$, at that redshift,
i.e.~sSFR$_\mathrm{first\,10\%}<0.3/t_z$.

\begin{figure}
\centering \includegraphics[width=85mm]{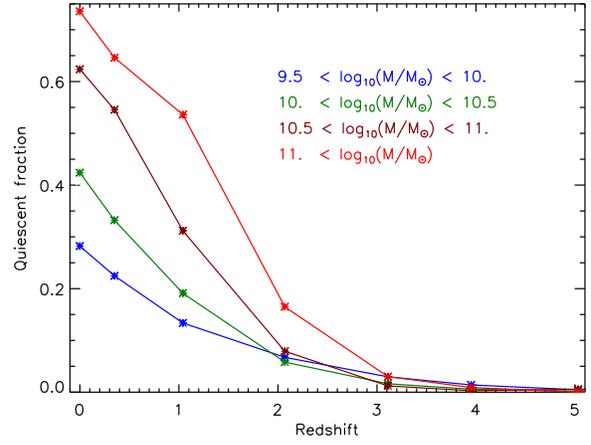}
\caption{The quiescent fraction of galaxies in the HWT14 model, in several
  different mass bins, and at several different redshifts, as shown.}
\label{fig:quiescent}
\end{figure}

\Fig{quiescent} shows the fraction of quiescent galaxies as a function of mass
and redshift.  At high redshift, there is some suppression of star formation in
dwarf galaxies, reflecting the strong feedback from supernova in the model.
However, only a small fraction of galaxies are affected and there is a much
greater growth of the passive population at redshifts below $z\approx3$.  
Once again, there is clear evidence of down-sizing in that more massive galaxies
start to become quiescent earlier than lower-mass ones.

\Tab{ssfr} lists several measures of star formation activity.
$S_\mathrm{0.5,all}$ is the median sSFR of the sample multiplied by the age of
the Universe at that redshift; likewise $S_\mathrm{0.5,sf}$ is the same thing,
but restricted to star forming galaxies.  The two begin to differ significantly
below a redshift of about 3 once the quiescent fraction begins to rise.  This
fraction is listed weighted both by galaxy number and by galaxy mass, from which
it can be seen that about 58 per cent of stars in the current-day Universe lie
in galaxies that are not actively star forming.

\begin{table}
  \caption{The median sSFRs, and the quiescent fraction
    of galaxies in the HWT14 model with mass exceeding $3\times10^9$\,\Msun.
    The columns are: redshift; age of the Universe in Gyr, $t$/Gyr; median specific
    SFR multiplied by the age of the Universe,
    $S_\mathrm{0.5,all}$; the same but restricted to star forming galaxies,
    $S_\mathrm{0.5,sf}$; quiescent fraction weighted by number,
    $Q_\mathrm{num}$; and quiescent fraction weighted by mass, $Q_\mathrm{mass}$.}
  \label{tab:ssfr}
\begin{center}
\begin{tabular}{llllll}
\hline
Redshift& $t$/Gyr& $S_\mathrm{0.5,all}$& $S_\mathrm{0.5,sf}$& $Q_\mathrm{num}$&
$Q_\mathrm{mass}$ \\
\hline
6.97&  0.77&  3.71&  3.71&  0.00&  0.00\\
5.03&  1.16&  3.17&  3.17&  0.00&  0.00\\
3.95&  1.56&  2.81&  2.82&  0.01&  0.01\\
3.11&  2.06&  2.50&  2.53&  0.03&  0.02\\
2.07&  3.17&  2.02&  2.09&  0.07&  0.08\\
1.04&  5.69&  1.43&  1.60&  0.18&  0.28\\
0.35&  9.79&  0.99&  1.31&  0.33&  0.48\\
0.00& 13.80&  0.69&  1.17&  0.41&  0.58\\
\hline
\end{tabular}
\end{center}
\end{table}

The medians listed in the table show that there has been a steady decline in
star formation activity in galaxies from a redshift of at least 7 right through
to the current day.  However, even as recently as $z\approx0.35$, most galaxies
were still forming stars at a rate that would more than double their mass within
the age of the Universe at that time.

\subsection{The cause of mass-dependent SFHs}
\label{sec:sfhmass}

\begin{figure}
\centering
\includegraphics[width=85mm]{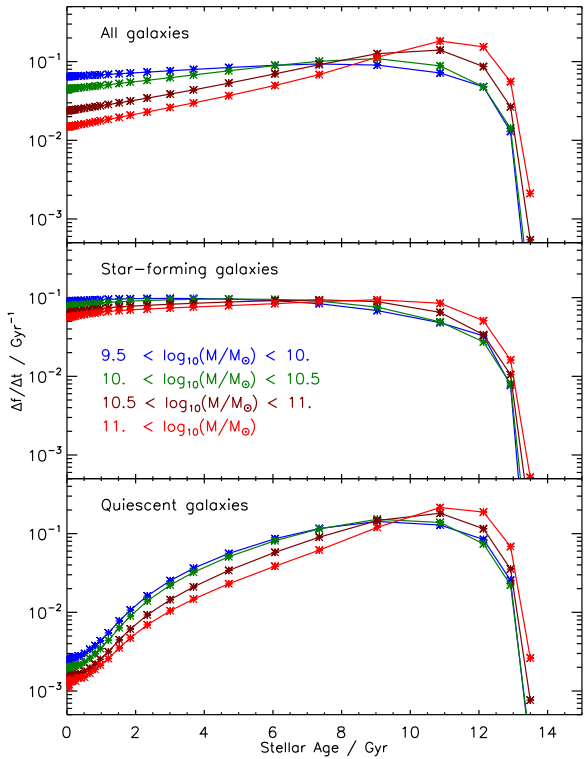}
\caption{The SFR as a function of time for the HWT14 model at
  $z=0$, split into 4 different mass bins, as shown.  The upper panel shows all
  galaxies in each mass-range; the middle panel shows star forming galaxies; and
  the lower panel panel shows quiescent galaxies, as defined in \Sec{ssfr}.}
\label{fig:sfhmass}
\end{figure}

The top panel of \Fig{sfhmass} shows the SFHs of galaxies at $z=0$ split into
four different mass bins. In order to better illustrate the onset of star
formation, which is poorly resolved at $z=0$ using our default number of bins,
for this section only we use $N_\mathrm{max}=4$ (see \Sec{sfhbin}), giving 29
time-bins at $z=0$.  Age downsizing is clearly visible with more massive
galaxies forming their stars earlier than lower-mass ones.

To better understand the cause of this, the middle and lower panels show the
same curves for star forming, and for quiescent galaxies, respectively.
Although the correspondence is not perfect, the agreement between the different
mass-bins is much tighter in the central panel than in the upper one.  That
strongly suggests that the SFHs of star forming galaxies are very similar,
independent of the mass of the galaxy, and that the primary driver of the
mass-dependence is the different evolution of the quiescent fraction.  Note that
we tried only a single definition of quiescence and it is likely that the
residual mass-dependence in the central panel could be reduced even further if
we optimised the definition for that purpose.

\subsection{Individual SFHs}
\label{sec:sfhind}

In this section, we restrict our attention to star forming galaxies.

\begin{figure}
\centering
\includegraphics[width=85mm]{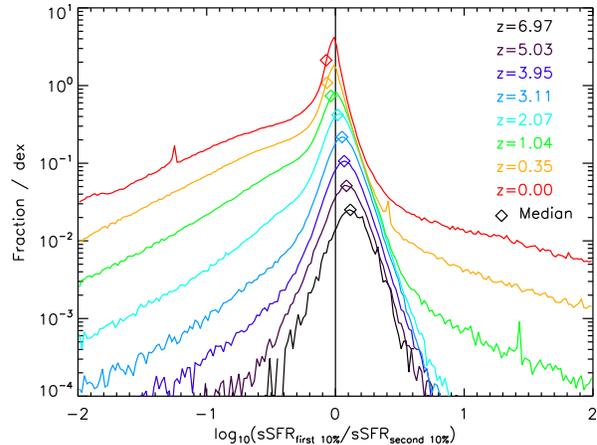}
\caption{The ratio of the SFR in the most recent ('first') 10
  per cent, to that in the previous ('second') 10 per cent, of the age of the
  Universe at that redshift, for galaxies with masses exceeding 10$^{9.5}$\Msun
  in the HWT14 model. The $y$-scale corresponds to the $z=0$ curve: the
  higher-redshift curves are offset by successive factors of two to space them
  out in the $y$-direction.  The diamond symbols show the median values.}
\label{fig:sfhind}
\end{figure}

Although the mean SFH is well-described by a simple functional form,
\Fig{sfhind} shows that individual galaxies have a wide variety of histories.
This figure shows histograms of the rate of decline of the SFR
measured by the ratio in two successive time-bins: the most recent 10 per cent,
and the next most recent 10 per cent, of the age of the Universe at that
redshift.  Galaxies to the left/right of the vertical line have
declining/increasing SFRs, respectively.

The distribution of ratios is shown in \Fig{sfhind} for a variety of redshifts.
There is a gradual shift from increasing to decreasing specific star formation
rates as the Universe ages.  When measured in this way, an equal balance between
increasing and decreasing SFRs is achieved somewhere between
redshifts 1 and 2.  At all times, however, there is a significant fraction of
galaxies lying in each of these populations: at $z=7$, four-fifths of galaxies
show an increasing SFR, and at $z=0$ three-quarters show a decreasing one.

\section{Conclusions}
\label{sec:conc}

In this paper, we have introduced the recording of star formation histories
(SFHs) in the \lgal\ \SAM.  At any given point in a galaxy's evolution, the mass
of recently-formed stars is recorded in bins of time resolution equal to that of
the timestep in the \SAM\ (1-2$\times10^7$\,yr).  These bins are gradually
merged together as the galaxy ages, such that older stars are grouped together
into larger bins.

We investigate the extent to which SFHs may be used to reconstruct stellar
spectra in post-processing; we compare our SFHs to those in the
publicly-available \Vespa\ catalogue extracted from SDSS-DR7 data; and we
investigate in our favoured \SAM\ \citep{HWT14} the evolution of SFHs
as a function of galaxy mass.  Our key results are as follows:
\begin{itemize}
\item Post-processing reconstruction of magnitudes in various observational
  bands gives good agreement with on-the-fly accumulation of luminosity,
  provided that $N_\mathrm{max}\geq2$ (which equates to $16$ bins at $z=0$).
  Quantitatively, the rms difference between raw and reconstructed
  magnitudes is less than 0.05 for all bands except the far-UV, for which it is
  0.29.
\item The \SAM{}s show reasonable qualitative agreement with the observed SFHs
  of the SDSS Main Galaxy Sample from the \Vespa\ catalogue, with the GWB11
  model fitting better for stars older than 1\,Gyr, and the HWT14 model fitting
  better for younger stars.
\item When divided the mass, both the observations and models show a trend for
  more massive galaxies to form their stars earlier and have lower current sSFRs
  than lower-mass galaxies.
\item When divided by colour, the agreement is poorer.  Both versions of the
  \SAM\ show much more extreme variation in SFH with colour than do observed
  galaxies from the \Vespa\ data base.  In the model, the SFHs of red and blue
  galaxies begin to differ as long ago as 5\,Gyr, compared to just 0.3\,Gyr for
  observed galaxies.  One possible explanation could be that real galaxies show
  repeated episodes of star formation that are not present in the models.  We
  note, however, that a more rigorous investigation of the data is required
  before drawing any definitive conclusions.
\item At $z\geq 1$ the mean SFR of all model galaxies with
  stellar mass greater than $3\times10^9$\Msun is well fitted by the formula
  d$f$/d$t\propto x^{1.5}e^{-x}$, where $x=(\ta-t)/2.0\,$Gyr.  Here $t$ is the
  lookback time and $\ta$ is the age of the galaxy.  At later times, the SFR
  declines less rapidly and a two-power model (that contains an extra parameter)
  is a better fit over the whole of cosmic history:
  d$f/$d$t\propto1/(x^{-1.39}+x^{1.33})$, where $x=(\ta-t)/3.0\,$Gyr.
\item Although star formation rates have been declining for more than half the
  history of the Universe, the typical (median) star forming galaxy today is still forming
  stars at a rate that will more than double its mass in a Hubble time.
\item We define a galaxy to be quiescent if it forms fewer than 3 per cent of
  its stars in the most recent 10 per cent of the age of the Universe, $t_z$ at that
  redshift, i.e.~sSFR$<0.3/t_z$. Then the quiescent fraction begins to increase
  rapidly below $z\sim 3$, reaching 41\,\% by number and 58\,\% by mass for
  galaxies with $M>10^9$\,\Msun\ at the current day.
\item Our model produces a small fraction of quiescent dwarf galaxies,
  $M<10^{10}$\,\Msun at all times, but the main effect that we see is consistent
  with downsizing in that more massive galaxies become quiescent first,
  followed by successively lower mass galaxies as the Universe ages.
\item When split by mass, and using a finer time resolution in the SFHs,
  downsizing is very clear to see, with the peak of the SFR shifting from a
  lookback time of about 11\,Gyr in the most massive galaxies (current-day mass
  greater than $10^{11}$\,\Msun) to less than 8\,Gyr in lower-mass systems
  ($3\times 10^9$--$10^{10}$\,\Msun).
\item When split into quiescent and a star forming populations, the differences
  between the mean SFHs of star forming galaxies of different mass is much reduced.
  Downsizing thus has its origin in an earlier transition from star forming to
  quiescent status in galaxies that are more massive at the current-day.
\item Although the mean SFHs are well defined, there is a huge dispersion in the
  SFHs of individual galaxies such that, even at the current day, many galaxies
  still have increasing SFRs.  At $z=7$, four-fifths of galaxies show an
  increasing SFR, and at $z=0$ three-quarters show a decreasing one; an equal
  balance between galaxies with increasing and decreasing SFRs
  is achieved somewhere between redshifts 1 and 2.
\end{itemize}

As can be seen from the above, one of the key drivers of galaxy evolution is the
rate at which star formation is quenched.  This is investigated in a companion
paper, Henriques et al. (in preparation), that undertakes a detailed comparison
with observations of the quenched fraction as a function of environment and
mass.  There it is shown that the HWT14 model does a much better job than
previous incarnations of the \lgal\ \SAM\ in terminating star formation in
massive galaxies, whilst allowing continued star formation in low-mass
satellites, though the quantitative agreement is still far from ideal.

An earlier paper, \citet{YHT13}, combined the SFHs with a
multicomponent model for stellar feedback to investigate the metallicity
evolution of galaxies.  This then enables us to construct metallicity histories
for galaxies along the lines of the SFHs presented in this paper.
Unfortunately, the observational data from VESPA is currently unable to
constrain the metallicity histories with any degree of certainty.

The low-resolution ($N_\mathrm{max}=2$) SFHs for the HWT14 \SAM, presented in
this paper, are publicly available to download from the Millennium
data base\footnote{\tt http://gavo.mpa-garching.mpg.de/MyMillennium/} and have
been used to reconstruct predicted fluxes in post-processing.  Higher resolution
catalogues are available from the authors upon request.

\section*{Acknowledgements}

The authors contributed in the following way to this paper.  SS undertook the
vast majority of the data analysis and produced a first draft of the paper and
figures.  PAT supervised SS, led the interpretation of the results, wrote the
bulk of the text.  BH led and wrote the section on post-processing of
magnitudes.  RT led the interpretation of the \Vespa results.  At different
stages each of the other authors (excepting GL) were responsible for discussion
of the results, shaping different parts of the paper, and helping to draft the
text.  GL provided the technical support to help integrate the post-processing
of galaxy magnitudes in the Munich GAVO repository (aka the Millennium
data base).

This paper draws upon data from the \Vespa\ data base, hosted by the ROE.  Much
of the data analysis was undertaken on the Cosma-4 supercomputer at Durham
and on the Apollo cluster at Sussex.

PAT, SJO and SW acknowledge support from the Science and Technology Facilities
Council (grant number ST/L000652/1).  RT acknowledges support from the Science
and Technology Facilities Council via an Ernest Rutherford Fellowship (grant
number ST/K004719/1).  GL was supported by Advanced Grant 246797 'GALFORMOD' from the European
Research Council and by the National Science Foundation under Grant no.
1261715.

\bibliographystyle{mn2e}
\bibliography{mn-jour,sfh}

\appendix
\section{The choice of \Vespa\ catalogue}
\label{app:vespa}

The \Vespa\ SDSS-7 catalogue \citep{TWH09} contains galaxies with a wide variety
of data quality, some showing reconstructed mass errors that are greater than
100\,\%.  In order not to bias the results, we include the whole sample in our
analysis.  We show average SFHs weighted by galaxy number,
rather than galaxy mass, so as to minimize the effect of the errors in the mass
reconstruction.  We have checked that restricting the analysis to the galaxies
with the best data quality does, in fact, lead to qualitatively similar SFHs.

\begin{figure}
\centering
\includegraphics[width=85mm]{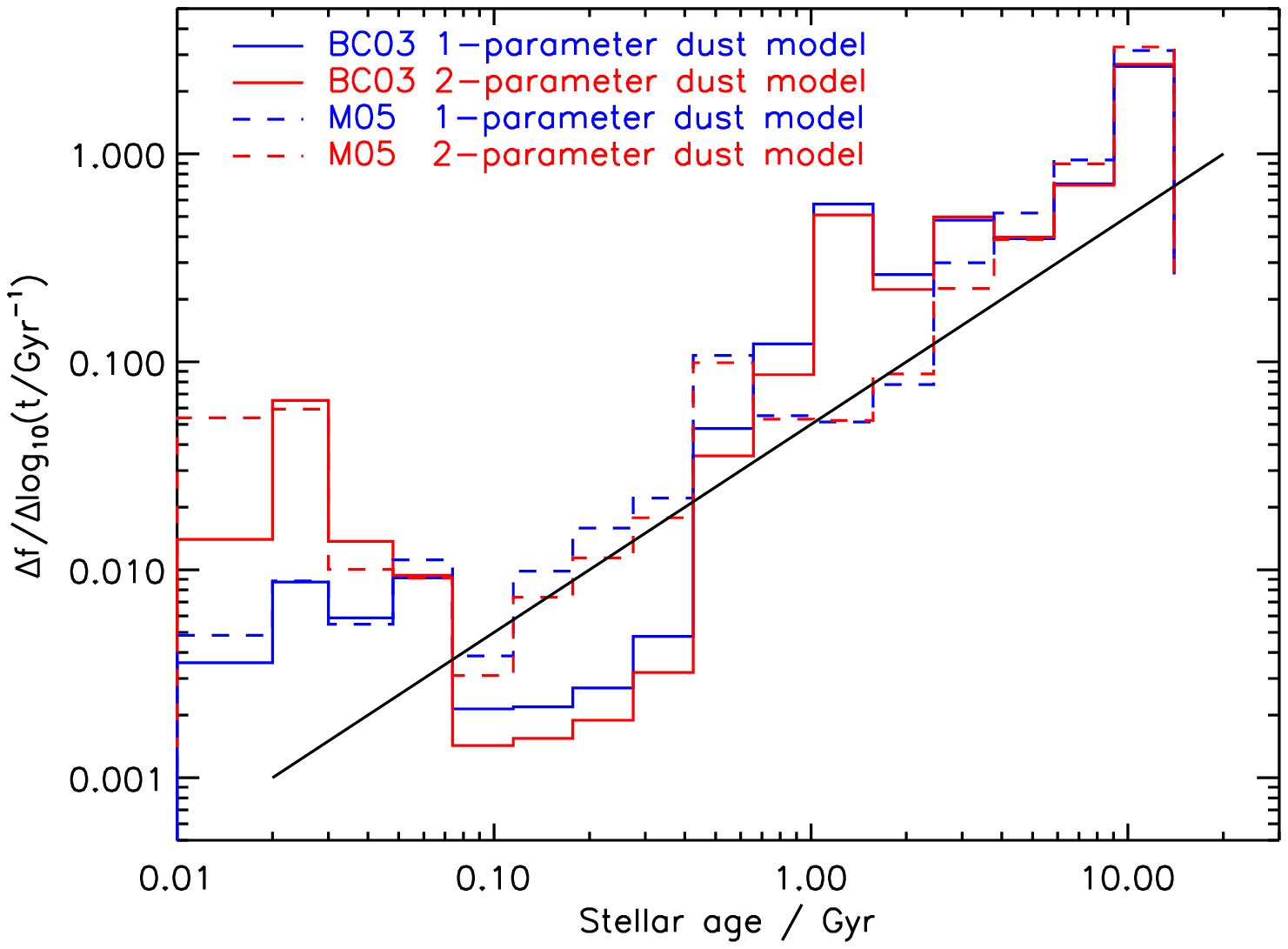}
\caption{The average SFH of VESPA galaxies obtained
  using the SEDs of BC03 (solid lines) and M05 (dashed lines). Blue
  lines show results for a one-parameter dust model whereas red shows a
  two-parameter dust model.  The solid, black line has a slope of unity,
  corresponding to a constant SFR.} 
\label{fig:vespaseddust}
\end{figure} 

The average SFHs of galaxies in this subsample, weighted by
galaxy number, are shown in \Fig{vespaseddust} for two different SEDs
(\citealt[hereafter BC03]{BrC03} and \citealt[hereafterM05]{Mar05}) and two
different dust models.  The one-component dust model is a uniform screen applied
to the whole stellar population; the two-component model adds in extra
absorption in front of young stars.

First note that the M05 models show a much smoother change in the SFR between
look-back times of 0.1-10\,Gyr than do those of BC03.  In such a large galaxy
sample, it is hard to think of a plausible reason for this and the \SAM{}s show
no such feature.  For that reason, we use the M05 results.

Both the BC03 and, to a lesser extent, the M05 results for the two-dust model show
a significant increase in SFR at ages less than 0.1\,Gyr.
Again, this seems implausible and suggests that there is not enough constraining
power in the data: the model has presumably confused dust obscured, young stars
with some older population, perhaps to explain some spectral feature that is not
well fitted by the SEDs.

Throughout the body of the paper, we use the VESPA results for M05 and a single
dust model.

\label{lastpage}

\end{document}